\documentclass{article}

\usepackage[english]{babel}

\usepackage[letterpaper,top=2cm,bottom=2cm,left=3cm,right=3cm,marginparwidth=1.75cm]{geometry}

\usepackage{amsmath}
\usepackage{graphicx}
\usepackage{float}
\usepackage[colorlinks=true, allcolors=blue]{hyperref}
\usepackage{wrapfig}
\usepackage{indentfirst}
\usepackage{enumitem}

\title{Uncoupling System and Environment Simulation Cells for Fast-Scaling Modeling of Complex Continuum Embeddings}
\author{G. Medrano, E. Bainglass, O. Andreussi}

\begin{document}
\maketitle

\begin{abstract}
Continuum solvation models are becoming increasingly relevant in condensed matter simulations, allowing to characterize materials interfaces in the presence of wet electrified environments at a reduced computational cost with respect to all atomistic simulations. However, some challenges with the implementation of these models in plane-wave simulation packages still persists, especially when the goal is to simulate complex and heterogeneous environments. Among these challenges is the computational cost associated with large heterogeneous environments, which in plane-wave simulations has a direct effect on the basis-set size and, as a result, on the cost of the electronic structure calculation. Moreover, the use of periodic simulation cells are not well-suited for modeling systems embedded in semi-infinite media, which is often the case in continuum solvation models. To address these challenges, we present the implementation of a double-cell formalism, in which the simulation cell used for the continuum environment is uncoupled from the one used for the electronic-structure simulation of the quantum-mechanical system. This allows for a larger simulation cell to be used for the environment, without significantly increasing computational time. In this work, we show how the double-cell formalism can be used as an effective PBC correction scheme for non-periodic and partially periodic systems. The accuracy of the double-cell formalism is tested using representative examples with different dimensionalities, both in vacuum and in a continuum dielectric environment. Fast convergence and good speedups are observed for all the simulation setups, provided the quantum-mechanical simulation cell is chosen to completely fit the electronic density of the system. 
\end{abstract}

\section{Introduction}

    Materials simulations have seen an incredible growth in recent years, thanks to the increasing power of computer hardware and simulation software, together with the development of new computational infrastructures to handle high-throughput simulations \cite{aiida1,aiida2,pymatgen,aflow,nomad}. Unbiased first-principles simulations based on density functional theory (DFT) represent the work-horse of most computational approaches to materials design \cite{marzari_ferretti_wolverton_2021}, with many research groups pushing to improve the accuracy of computational predictions and the extend the scope of first-principles simulations to more challenging systems and properties. 

    Traditional DFT simulations of materials focus on bulk periodic crystalline structures. For this reason, most condensed-matter and materials simulation packages rely on basis sets composed of periodic functions (e.g. plane-waves, PW) which uniformly occupy the simulation cell. This approach provides a systematically improvable basis set, whose accuracy can be tuned with a single parameter, usually associated with the kinetic energy of the corresponding plane-wave. PWs also provide easy access to the solution of the core electrostatic equations involved in first-principles simulations, with the Poisson equation in vacuum being readily solved in reciprocal space. Fast Fourier transforms (FFTs)\cite{fftw} have proven instrumental to the wide-spread success and fast scaling of PW-based DFT packages, by providing real-to-reciprocal space mapping, and vice-versa, at a cost proportional to $N_{pw}\log(N_{pw})$, with $N_{pw}$ being the number of PWs in the simulation. However, what is advantageous for bulk crystalline materials can be a source of artifacts when non-periodic or partially periodic systems need to be characterized. Traditionally, materials interfaces are simulated using the slab approach, in which a few-atoms thick slab of the material is described in a 3-dimensional periodically repeated simulation cell. By increasing the thickness of the slab and by increasing the cell size in the direction orthogonal to the material interface, the periodic replicas of the slab are effectively uncoupled from each other and the results are considered to be representative of the surface of a semi-infinite bulk material. A similar strategy can be devised for systems that are only periodic in one dimension, such as nano-wires or nano-tubes, or that are non-periodic. While adjusting the cell size can provide an easy solution to periodic boundary conditions (PBC) artifacts, the number of PWs grows linearly with the cell volume, leading to a polynomial increase in the cost of the corresponding DFT calculation. For this reason, multiple alternative strategies to correct for PBC artifacts have been proposed in the past years, including real-space and reciprocal-space approaches. Most of these strategies are focused on simulations of two-dimensional and zero-dimensional (isolated) systems, while one-dimensional correction schemes are usually less wide-spread.  

    Unfortunately, as the typical scaling of DFT simulations depend varies as $N^3$, where $N$ is the number of electrons in the system, complex and large systems with more than a few hundreds of electrons are still out of reach for systematic DFT studies. To overcome this limitation, a variety of hierarchical approaches \cite{continuum-review,QMMM-rev} and/or divide and conquer\cite{eqe,sub-sys1,gga-sub-sys} strategies have been developed in the literature. Without sacrificing the important atomistic characteristics of the relevant part of the process under investigation (the system), these methods take advantage of different, possibly simplified, models to handle the more macroscopic or complex part of the process (the environment), thus allowing a significant reduction of the computational cost.
    
    In particular, implicit solvation models based on continuum embedding media have proven to provide a reasonable qualitative accuracy, while significantly reducing the need for statistical sampling of disordered configurations \cite{orozco_luque_chem-revs,cramer_truhlar_chem-revs,tomasi_persico_chem-revs,tomasi_mennucci_cammi_chem-revs}. Indeed, removing the atomistic details of solvent molecules not only reduces the number of electrons involved in the DFT simulation, but it allows to avoid the use of molecular dynamics (MD) to sample the configurations of the molecules of the liquid. Continuum methods in computational chemistry usually involve more or less empirical definitions of the different contributions to solvation free energy, which are then parameterized on experimental databases generated from solubility data. Methods such as the Polarizable Continuum Model (PCM)\cite{PCM} of Tomasi and co-workers have shown a good success in predicting ground state and response properties of molecules in solution. More recently, similar approaches have been translated into condensed-matter simulation packages in order to study wet, possibly electrified, interfaces of materials\cite{continuum-review}. When compared to the more widespread continuum solvation models in the computational chemistry community, the models implemented in periodic PW-based simulation packages rely on a smoothly-varying boundary between the quantum-mechanical system and the continuum environment, instead of a sharp two-dimensional interface. Definition of the interface based off the electronic density of the solute\cite{SCCS} or its atomic positions\cite{SSCS} have been developed and show similar accuracy, when properly parameterized. While computationally more demanding, the use of smooth interfaces allows continuum models in condensed matter to more seamlessly introduce non-local corrections \cite{SSSA,field-aware} to avoid some of the typical artifacts of these solvation methods. 

    Beyond the use of continuum solvation to model bulk neutral solutions, extensions to handle liquid crystals and diluted ionic solutions were implemented exploiting the integral equation formalism (IEF) of PCM\cite{PCM-IE}. Still following the same strategy, PCM was extended to the study of a single two-dimensional interface between different dielectric media (liquid-air, liquid-liquid)\cite{diffuse_layers}. Corni et al. extended the use of the polarizable dielectric model to handle the electrostatic effects of metal surfaces and nanoparticles on nearby molecular dyes, thus allowing a more quantum-mechanical characterization of surface-enhanced spectroscopies\cite{corni_tomasi_2001,andreussi_corni_mennucci_tomasi_2004,andreussi_biancardi_corni_mennucci_2013}.
    
    Similar to the PCM literature, more complex and heterogeneous environments can be introduced in condensed-matter simulations. In particular, a significant focus has been devoted to study electrolyte solutions\cite{diffuse_layers,GCE-DFT,sundararaman_schwarz_2017,jinnouchi_anderson_2008,nattino_truscott_marzari_andreussi_2019,jdftx}, so as to unlock the use of these models for simulations of electrified interfaces. Going beyond neutral solutions and electrolyte distributions, Campbell et al. \cite{quinn-paper,quinn-erratum} exploited a continuum approach to model the charge reorganization in a semiconductor substrate. 
    
    As the definition of the continuum environment in smooth-interface solvation models follows a function defined everywhere in the simulation cell, it is relatively straightforward to introduce more complex and heterogeneous environments, e.g. by introducing multiple continuum media, each with its own boundaries and physical properties (e.g. dielectric constant or surface tension). These media can be used to model substrate effects on the properties of overlaying materials or molecular compounds. Following this idea, Bononi and collaborators\cite{fernanda,fernanda2} exploited the possibility to introduce flat two-dimensional dielectric regions in the simulation cell in order to characterize in a computationally inexpensive way the effect of an ice substrate on the absorption properties and photo-degradation of molecular dyes. 

    Despite the flexibility of continuum embedding approaches in condensed matter  environments, PW-based simulations can pose significant challenges to these class of multiscale methods. The use of periodic simulation cells is intrinsically unfit to model systems that are embedded in a semi-infinite medium, i.e. most solvated systems would typically require non-periodic simulation cells in the directions where the continuum solvent resides. Moreover, the fact that the simulation cell size is directly related to the basis-set size $N_{pw}$ makes simulating large heterogeneous environments computationally more challenging. Extensions of PBC correction schemes have been developed to account for continuum dielectric media in simulations of isolated (0D) and slab (2D) systems\cite{andreussi_marzari_2014}. In both continuum electrolyte and semiconductor models, long-range charge reorganizations are handled implicitly, by relying on analytical solutions to the electrostatic problems and/or appropriate boundary conditions. However, simulating the effects of nano-sized environments, such as plasmonic nanoparticles, micelles, liquid nanodroplets, would require to use simulations cells, and thus basis-sets, that are beyond reach of standard computational resources. 
    
    The straightforward solution to the aforementioned problem is to uncouple the simulation cell exploited for the continuum environment from the one used for the DFT simulation of the quantum-mechanical system. The use of a large simulation cell for the environment would not cause significant limitations to the simulation time: the time-consuming part of the continuum environment calculation is related to computing the electrostatic potentials, by solving modified and generalized forms of the Poisson or Poisson-Boltzmann equations. Most of the algorithms implemented rely on iterative strategies in which each iteration involve the solution of a simple Poisson equation, which can be performed via FFTs in a time almost linear in the cell size. 
    By keeping the simualtion cell of the DFT calculation small, the largest overhead in computational time is avoided. While ideal in theory, in practice this double-cell formalism requires to fully uncouple the two simulation methods and to carefully design the mapping between the two simulation cells. 
    
    While designed to account for large heterogenous environments, the double-cell formalism can also be used for non-periodic and partially periodic systems as an effective PBC correction scheme. By increasing the length of the lattice vectors going along the non-periodic directions in the environment cell, electrostatics can be fully converged without affecting the basis-set size and cost of the quantum-mechanical calculation. This general strategy works equally well for isolated (0D), slab (2D), but also one-dimensional (1D) systems. In this work, the details of the double-cell methodology are presented and the approach is tested as a PBC correction scheme for some representative examples with different dimensionalities, both in vacuum and in a continuum dielectric environment. Along with this new methodology and in order to test the accuracy and the speed-up of the double-cell formalism for PBC corrections, the  auxiliary-function countercharge corrections library (LIBAFCC) of Dabo et al.\cite{afc90} was coupled to the continuum embedding software, giving access to an alternative exact correction scheme for all types of partially periodic systems.

    The paper is organized as follows: in Section \ref{sec:theory} we review the main details of the implementation of the double cell formalism, as well as the coupling of the auxiliary-function countercharge corrections library (LIBAFCC)\cite{afc90}; in Section \ref{sec:comp} the comptuational details of the DFT calculations as well as how the double cell calculations are reported. Eventually, in Section \ref{sec:results}  we report a detailed benchmark of the accuracy and performance of the double cell approach for modelling partially periodic systems, including 0D, 1D, and 2D systems, in vacuum and in a uniform dielectric medium.

\section{Methods}\label{sec:theory}

    \subsection{Double-Cell Formalism}
        
    \begin{figure}[h]
        \begin{center}
            \centerline{\includegraphics[scale=0.75]{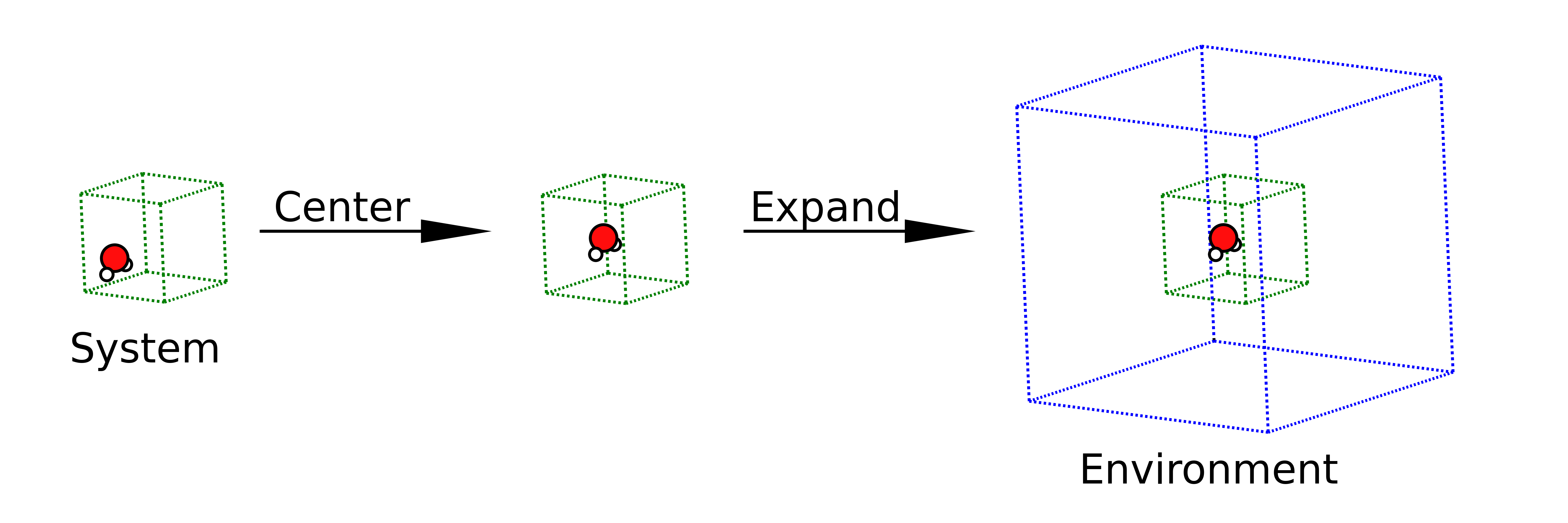}}
        \end{center}
        \caption{A schematic view of the double cell formalism with an isolated water molecule. The green cell is the system cell where all the PWs are introduced while the blue cell is the environment cell that are used for PBC correction.}
        \label{fig:dblcell}
    \end{figure}

    The design of a double-cell algorithm is strongly dependent on the numerical details that are associated with the calling program (e.g. the PW DFT simulation package), the continuum embedding library, and with their coupling. In particular, the key aspects that affect the implementation are:    
    \begin{enumerate}[leftmargin=3\parindent]
        \item which scalar fields need to be passed from one cell to the other
        \item how the scalar fields are stored and parallelized in the two cells
        \item the way the gridpoints of the two cells are aligned
    \end{enumerate}

    In the following we will refer to the specific implementation involving Quantum Espresso as the DFT driver and Environ as the continuum embedding library. In order to generate the continuum boundary in Environ, information on the electronic density of the system and/or its atomic positions are required. The same information is also needed for the calculation of the electrostatic potential, i.e. the position and valence of the ions as well as the full electronic density of the system are required inputs for an Environ calculation. In alternative polarizable dielectric models, such as PCM, the electrostatic potential of the system in vacuum is passed to the continuum embedding module, to be used to solve for the polarization charge on the environment. However, passing the full information of the charge of the solute allows to have a consistent definition of the electrostatic potential, e.g. when smeared ionic charges are used to account for core-electrons. In the case of the double-cell formalism, passing the full charge density of the solute also allows to seamlessly adapt the calculation to a different simulation cell. In the presented implementation, the key scalar quantity that needs to be passed from the system cell into the environment cell is thus the electronic density of the system. 
    
    The results of the continuum embedding calculations are usually reported in terms of corrections to the corresponding quantities computed in vacuum and periodic boundary conditions by the host DFT program. In particular, corrections to the total energy, interatomic forces, and Kohn-Sham potential are the key outputs of Environ calculations. Of these quantities, only the latter needs to be mapped from the environment cell into the system cell. 

    Scalar fields in a 3D simulation cell are discretized in PW simulation packages in terms of their values on a 3D structured grid, whose spacing is inversely related to the PW cutoff specified in the simulation input. However, the values on the 3D grid are stored by Quantum-Espresso as one-dimensional arrays, by scanning grid-points sequentially along the three cell axis. As part of the hierarchy of parallelization schemes adopted by Quantum-Espresso, parallelization on the real-space simulation cell is implemented by dividing it into slices along the third axis or into sticks along the second and third axes, depending on the total number of available processors. This allows to perform operations that only depend on local values of the scalar field in a fully parallelized way, while still requiring gathering and scattering of the whole simulation data for real to reciprocal-space transformations (FFTs and inverse FFTs). Owing to the origin of Environ as an internal plugin of Quantum Espresso, the same strategies and the same numerical libraries are exploited by Environ for its internal calculations. 
    
    In designing the coupling between system and environment simulation cell, it is crucial to realize that the quantum-mechanical system lives in a fully periodic simulation cell: in the DFT calculation the system energy and properties are not affected by arbitrary translations of the system's degrees of freedom. In particular, it is possible for a system to have some of its atoms or part of its electronic density on opposite sides of the simulation cell, as the minimum image convention would still give rise to a well-defined fully connected system. However, when expanding the system cell into the environment cell, care must be taken to ensure that a fully connected system is passed to continuum embedding module and that the environment is properly defined with respect to a meaningful center of the system. Thus, for the mapping of scalar fields between the two simulation cells, a two step process was implemented as visualized in Figure \ref{fig:dblcell}: first a centering of the system and second an expansion of the environment cell along the designated directions (x,y,z).

    The first step is accomplished by computing the center of mass of the system within its original simulation cell, using the minimum image convention to account for potential splits in neighbouring cells. We then introduce a new cell with the same cell parameters and the same grid spacing as the system cell, but with an origin shifted from the Cartesian origin so as to place the system's center of mass near its central gridpoint. A one-to-one mapping of gridpoints, i.e. a change in the integer index of the stored one-dimensional arrays containing the scalar fields, can be computed on the fly without loss of information and without affecting the real-space parallelization scheme. 

    The second step of the double-cell algorithm involves the expansion the system cell in the different directions in order to generate a larger environment cell. While arbitrary expansions could in principle be considered, they would usually result in a environment grid spacing and gridpoint positions that would not match the initial ones on the system cell. This numerical mismatch could be easily addressed by interpolation algorithms, which would add an additional layer in the mapping between the two cells. However, in order to keep the process as simple and robust as possible, the current implementation of the double-cell formalism restricts the cell expansion to integer multiples of the initial cell. An expansion vector $\mathbf{n}=\left(n_1,n_2,n3\right)$ is defined in the input of the environment module, with $n_i$ representing the number of fictitious replicas added on both sides of the re-centered system cell along its $i$-th axis. The combination of these two steps creates the desired environment cell, with lattice vectors $2\mathbf{n}+1$ times as large as the system cell, with the center of mass of the system located near its central gridpoint. 

    \subsection{Auxiliary-Function Correction}


    When simulating materials using DFT, electron interactions can be hindered by computational singularities, becoming apparent when looking at partially periodic systems, arising from the reciprocal-space summation of interaction contributions. Most simulation software set these divergences to zero, introducing negligible errors that approach zero as the size of the computational supercell increases. Real-space electrostatic corrections have been introduced that can remove these singularities although they can become costly due to these corrections requiring different interaction potentials per self-consistent iteration.

    Because of this additional computational cost, reciprocal-space countercharge corrections have been studied in the form of auxiliary-function techniques which are the basis of the LIBAFCC\cite{afc90}. These techniques have an added flexibility with the choice of auxiliary functions which all have different definitions as well as capabilities in removing PBC errors. The LIBAFCC library implements exact point-charge reciprocal-space auxiliary functions for all types of partially periodic systems with the 0D correction giving the same formulas as the Martyna-Tuckerman\cite{MT} (MT) correction scheme. 

    Following the similarity of the AFCC with the MT approach, we coupled the library of Li and Dabo to a development version of Environ. The point-charge correction as computed by the library is mapped in the real-space grid, centered on the origin. Fast Fourier transforms allow to convert the correction to reciprocal space and to store it, so that it can be used to complement the standard vacuum FFT Poisson solver at no extra cost. 

\section{Computational Details}\label{sec:comp}


    All calculations were performed using Quantum Espresso (QE) v7.1\cite{qe1,qe2} compiled with the Environ library for continuum embedding effects \cite{SCCS,qe1,environ3}. While the double-cell implementation has been officially released in Environ 3.0, for the results reported ion the following we used a local development version of the library that includes the coupling with the AFCC library for periodic boundary corrections. 
    
    All reported benchmarks, in vacuum and in dielectric environments, involve a single self-consistent field (SCF) optimization of the electronic density, starting from a random initial guess. In an effort to keep benchmarks as consistent as possible, we did not include geometry optimization calculations in our results. However, convergence of the inter-atomic forces with the newly implemented algorithms was thoroughly tested, showing an accuracy consistent with fully periodic simulations and with other PBC correction schemes. 
        
    \begin{figure}
        \begin{center}
            \centerline{\includegraphics[scale=0.75]{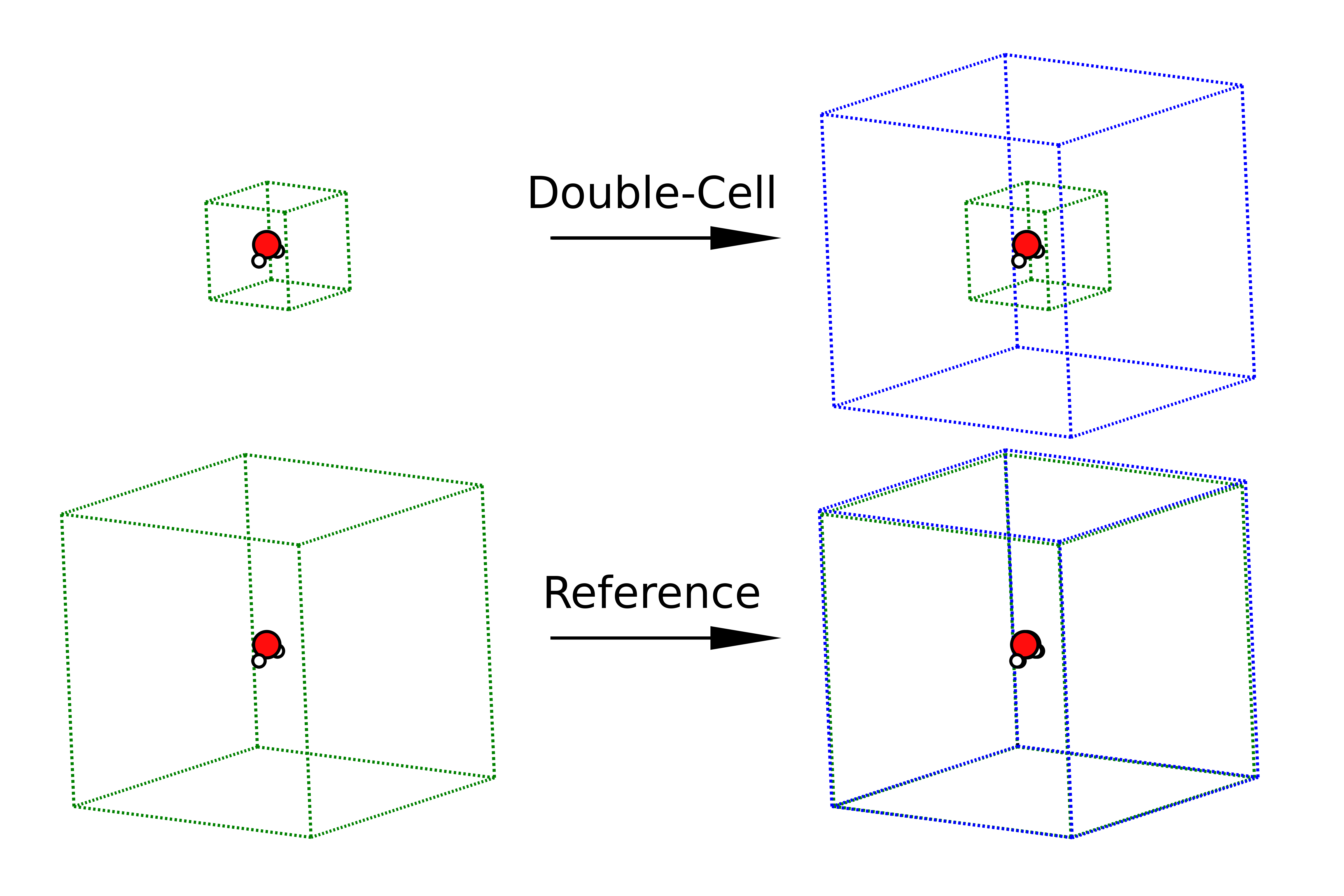}}
        \end{center}
        \caption{Graphic representation of the two sets of calculations adopted to investigate convergence of the developed formalism. The double-cell calculations involve a small system cell (green box) embedded in an expanded environment cell (blue box). Reference calculations adopt the same expanded cell for both the system (DFT) and the environment (continuum electrostatic) calculations.}
        \label{fig:calcs}
    \end{figure}

    The 0D and 2D simulations are based off the Environ's examples distributed with the library. In particular, the isolated system considered in this study is an acetamine cation, while a thin Pt (111) slab with an adsorbed carbon monoxide molecule was used for testing 2D systems. A BN nanoribbon was chosen as benchmark for the 1D systems. The selected systems are known to display more pronounced PBC artifacts. However, it is important to stress that the reported analysis is meant to focus on the consistency of the results with alternative PBC correction schemes and on the overall computational performance. For this reason, the physical approximations and the simulation details that are not relevant for the convergence of the electrostatic energy have been selected according to the most wide-spread choices in the literature. All the calculations were performed using the Perdue-Burke-Ernzerhof  generalized gradient approximation (GGA) density functional\cite{PBE}. For the 0D and 2D systems, the standard ultra-soft psuedopotentials distributed on the QE website were adopted. For the 1D system, the pseudopotentials from the SSSP efficiency library were selected\cite{SSSP1,SSSP2}. The Simulations were performed by only sampling the gamma point sampling of the Brillouin zone. 

    A dielectric embedding environment was modelled using the self-consistent interface function of Andreussi et al\cite{SCCS}. A homogeneous static dielectric permittivity of 100 was selected for the bulk of the environment. The preconditioned conjugate gradient (PCG) approach of Fisicaro et al \cite{PCG} was used for the convergence of the generalized Poisson equation, with a convergence threshold on the computed electrostatic potential of $5.0e-13$ Ry \cite{SCCS,qe1}.

    For the acetamine cation (0D system) the wave function cutoff and density cutoff  were chosen to be 30 Ry and 300 Ry, respectively, with a estimated convergence threshold on the total energy of $5.0e-6$ Ry. A simple cubic simulation cell was adopted, in order to allow the use of a parabolic correction to PBCs \cite{para1,para2}. For this isolated system, reference simulations were also performed using the Martyna-Tuckerman reciprocal space correction \cite{MT}, both in vacuum and in the dielectric medium.
    
    For the 2D system, an orthorhombic cell was used with the cell parameter in the z-direction being varied to evaluate the interaction with periodic replicas. The wave function cutoff and density cutoff were set to 35 Ry and 300 Ry, respectively, with an estimated convergence threshold on the total energy of $1.0e-6$ Ry. Marzari-Vanderbilt\cite{MV} smearing of the band occupations was adopted with a spread value of 0.03 Ry. A parabolic correction scheme, as implemented in the Environ library, was adopted for simulations in vacuum and in the dielectric medium.
    
    Finally, for the 1D system, an ideal boron-nitride (BN) 2D surface, as obtained from the aflowlib\cite{aflow} repository, was used to create a 1D ribbon. The wave function and density cutoffs were set to 60 Ry and 480 Ry, respectively, with an estimated convergence threshold on the total energy of $1.0e-7$ Ry.  A Gaussian smearing with a spread value of 0.02 Ry was adopted, together with a local Tomas-Fermi mixing and a mixing parameter of 0.7. Although the parabolic correction scheme implemented in Environ is only compatible with 0D and 2D systems, a 2D correction with a large simulation cell in the direction along the plane of the ribbon was also exploited as reference for the 1D system. 

    In order to demonstrate the accuracy and performance of the double-cell formalism, two DFT calculations were done for each unit cell size, as schematized in Figure \ref{fig:calcs}. The first calculation is performed using double-cell approach, i.e. having the environment cell defined as an integer expansion of the system cell in at least one direction. A reference calculation is then performed by imposing the system and environment cells to be of the same size and to match the environment cell of the first calculation. As an example, for a simulation of an isolated (0D) system with a cubic cell of size 10 Bohr, a double-cell calculation with one replica in every direction, $\mathbf{n}=\left(1,1,1\right)$, would involve a cubic environment cell with a 30 Bohr side length. The reference calculation would be performed with identical system and environment cells, both set to a size of 30 Bohr. In the following section, the energy difference between the double-cell calculation and the reference are reported together with the speed up in calculation time. 

\section{Results}\label{sec:results}

\subsection{Isolated Systems (0D)}\label{sec:0D}
    
    \begin{wrapfigure}{r}{0.5\textwidth}
        \begin{center}
            \includegraphics[scale=0.125]{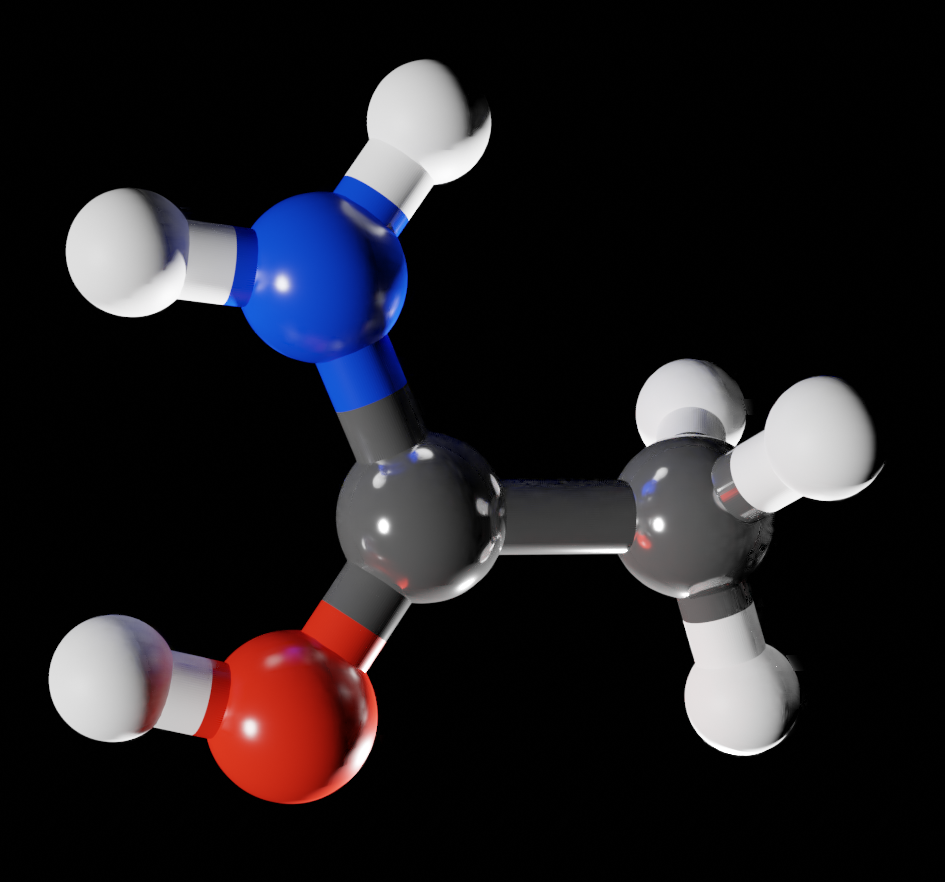}
        \end{center}
        \caption{The acetamine cation used for testing the double-cell formalism in isolated systems.}
        \label{fig:acetamine}
    \end{wrapfigure}

    As discussed above, one of the uses of the double-cell formalism  is to reduce PBC artifacts in isolated systems, thus overcoming one of the main sources of artifacts of plane-wave based codes. The system selected for testing the accuracy and performance of the double-cell strategy is an acetamine cation, visualized in Figure \ref{fig:acetamine}. To simplify the handling of PBC artifacts and to allow the use of the parabolic point-counter-charge correction, a cubic cell is used to study this system. 
    
\begin{figure}
        \begin{center}
            \centerline{\includegraphics[scale=0.29]{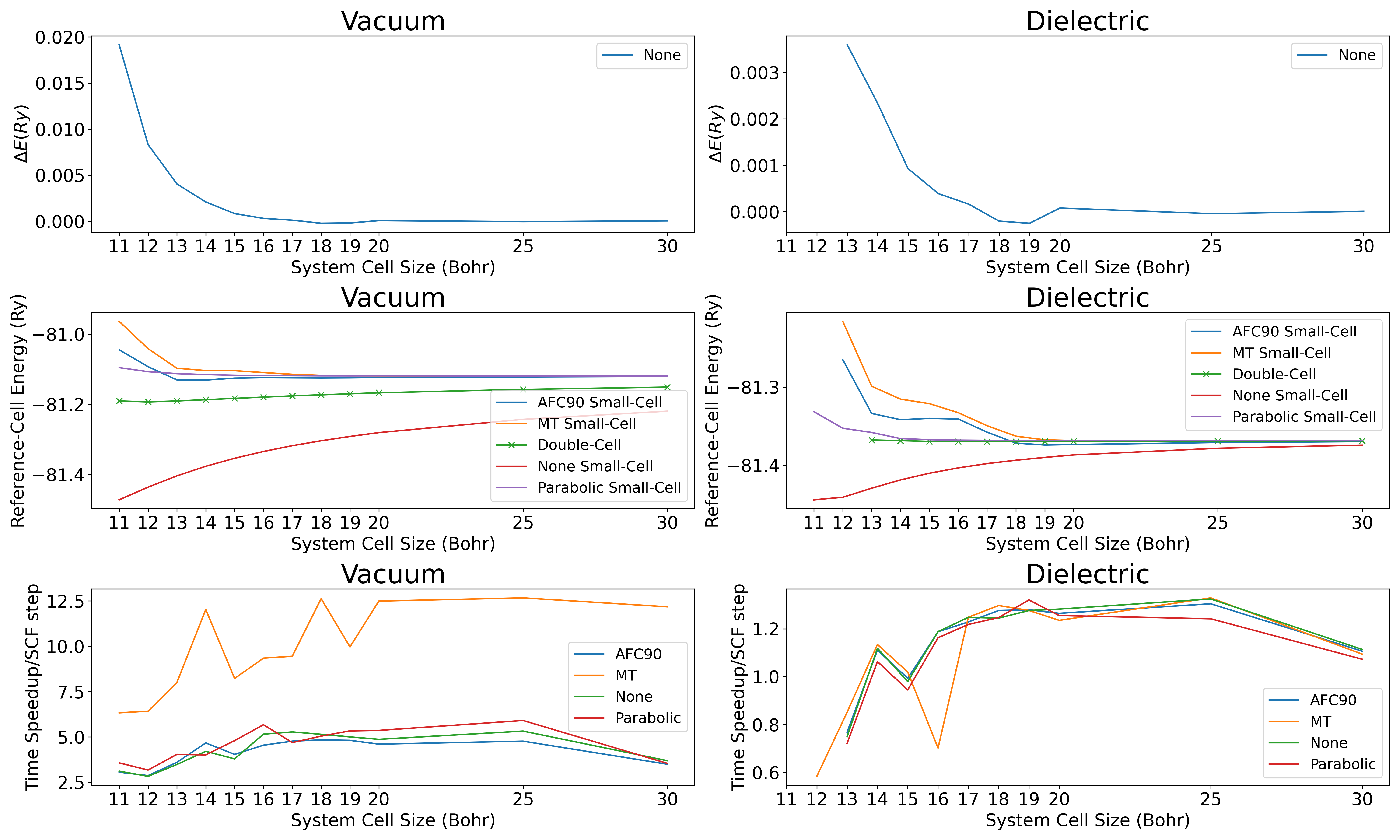}}
        \end{center}
        \caption{The top row plots show the energy difference, in Ry, between an SCF calculation using the double cell formalism and the corresponding reference calculation (as visualized in Figure 2). In the middle row, the convergence of different PBC schemes as a function of DFT system cell size is reported. The bottom row plots show the time speedup between these simulations. All reported tests were performed for systems in vacuum (left column) and in a continuum dielectric medium (right column). }
        \label{fig:0d-res}
    \end{figure}

    In the top panels of Figure \ref{fig:0d-res} the changes in total energy with respect to the reference are reported, for both simulations in vacuum and in a dielectric environment. The considered molecule has overall size of about 8.0 Bohr. However, from the convergence of the double-cell calculations with respect to the reference it is clear that system cells smaller than 15 Bohr present artifacts with respect to the same calculation in the reference cell. As the electrostatic calculations in the double-cell and the reference are the same, these small difference must be due to the fact that the double-cell mapping is affecting the DFT convergence. In fact, inspection of the density of the molecule reveals that the difference is due to the non-negligible spilling of the electronic density of the molecule beyond the cell boundaries. While in a periodic cell this still corresponds to a connected smooth density, the mapping step of the double-cell formalism breaks the spilling density apart. This result suggests a general rule for the identification of the minimal system cell in a double-cell calculation, which needs to be about 10.0 Bohr larger than the system size.  
    
    When looking at the behaviour of the double-cell energy as a function of system size, it is clear that the double-cell results are fully converged for cells larger than the identified minimum size. This is consistent with all the other PBC correction schemes investigated, which also show converged electrostatic for cell sizes larger than 15 Bohr in vacuum, and 20 Bohr in a dielectric medium. In the bottom panels of Figure \ref{fig:0d-res}  we report the speedup per SCF cycle of the double-cell formalism when compared with the reference calculations. This speedup accounts for the relative weight of the FFT-based electrostatic calculation and the cost of inverting the DFT Hamiltonian in a single SCF step. The reported results show an average five-fold decrease in computational time when the system cell is allowed to be smaller than the environment cell. For the case of dielectric embedding, iterating over the solution of the Generalized Poisson problem in the dielectric environment increases the relative costs of the electrostatic calculation with respect to the SCF step, thus producing a significantly smaller speedup. 

    While the results in this section aim to provide a comparison between alternative PBC schemes, it is important to stress that the double-cell formalism can also be used in combination with any other PBC corrections scheme reported above: the electrostatic calculation in the expanded environment cell can be corrected with a real-space term (parabolic correction) or with any reciprocal-space scheme (Martyna-Tuckerman or AFC) already implemented in Environ. Those results are not reported in this section because the simulations are already fully converged without the need for additional corrections. 
    
\subsection{Slab (2D) System}\label{sec:2D}

\begin{wrapfigure}{r}{0.5\textwidth}
        \begin{center}
            \includegraphics[scale=0.125]{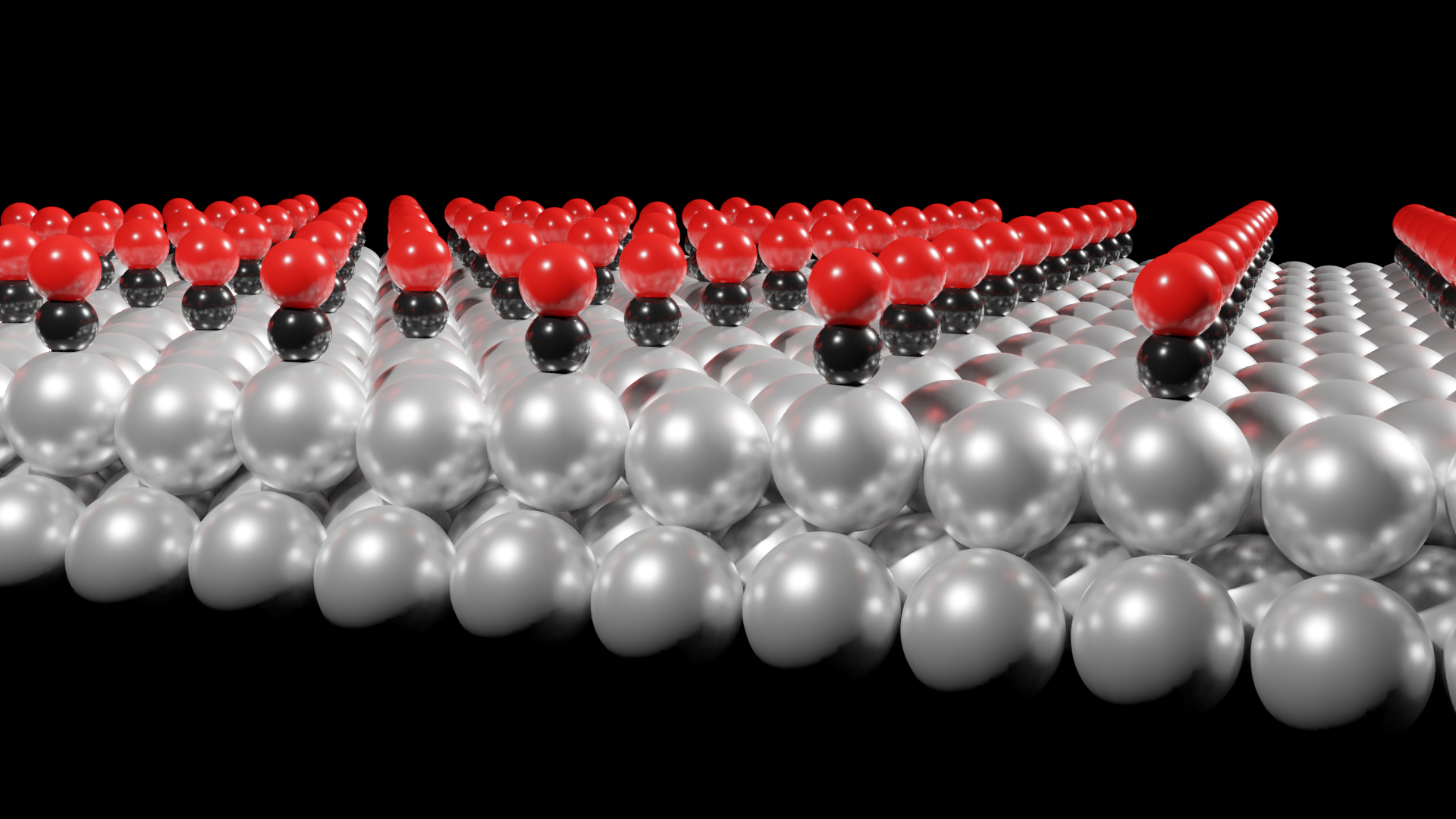}
        \end{center}
        \caption{Visualization of the system used for testing the double-cell algorithm in two dimensions: a Pt (111) surface with two layers of Pt atoms in the presence of a CO molecule adsorbed in the atop position.}
        \label{fig:ptco}
    \end{wrapfigure}

\begin{figure}[H]
        \begin{center}
            \centerline{\includegraphics[scale=0.29]{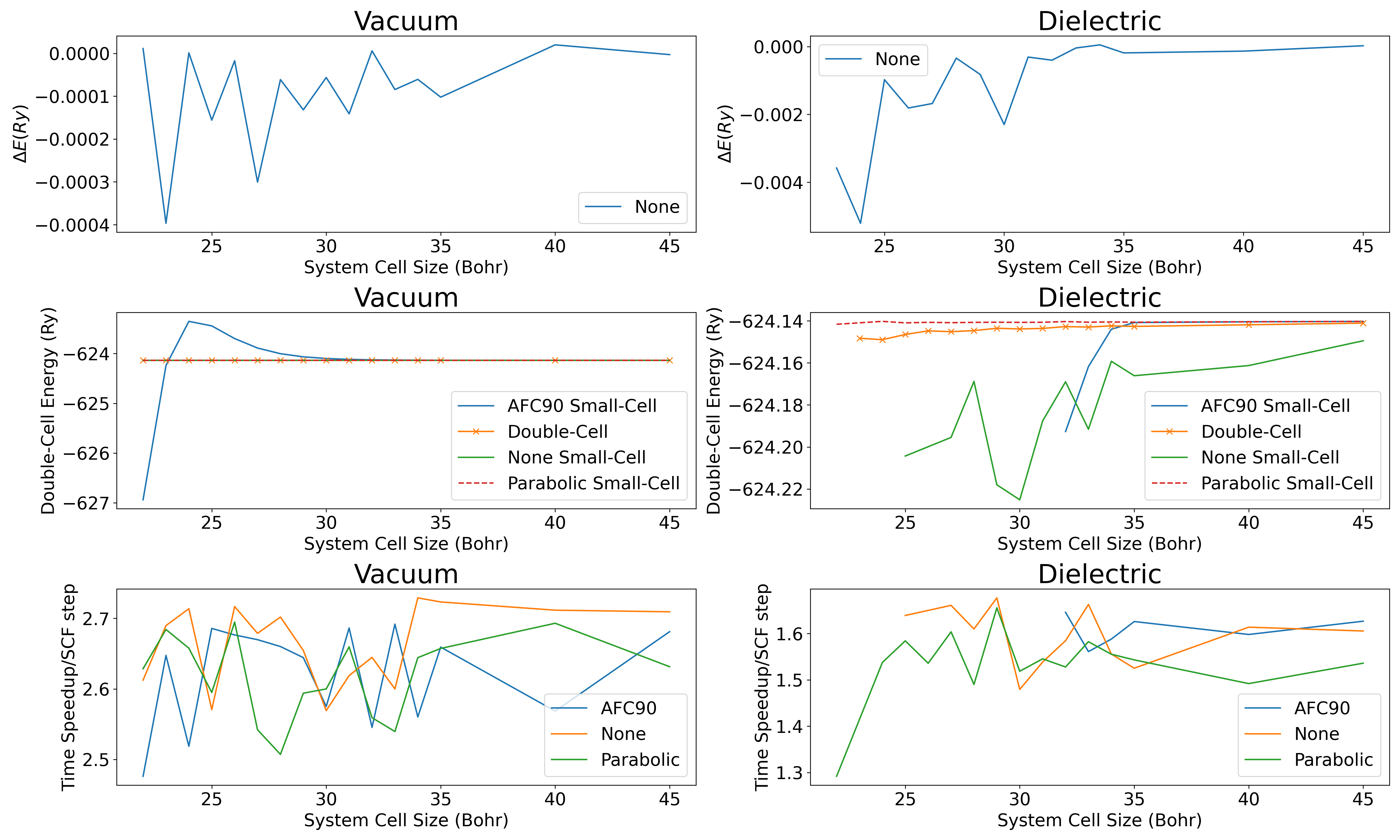}}
        \end{center}
        \caption{The top row plots show the energy difference, in Ry, between an SCF calculation using the double cell formalism and the corresponding reference calculation (as visualized in Figure 2). In the middle row, the convergence of different PBC schemes as a function of DFT system cell size is reported. The bottom row plots show the time speedup between these simulations. All reported tests were performed for systems in vacuum (left column) and in a continuum dielectric medium (right column).}
        \label{fig:2d-res}
    \end{figure}

    To study 2D systems, systems that should be periodic in two dimensions, we used a bi-layer of platinum atoms in the (111) orientation and in the presence a CO molecule in the atop position, as visualized in Figure \ref{fig:ptco}. For this system, the slab is included in an orthorhombic cell,  oriented in the x-y plane and with the non-periodic axis along third direction. For the double-cell calculations, the system cell is expanded only in the z direction, while the system and environment cells are the same in the in-plane directions. Convergence tests are reported in Figure \ref{fig:2d-res} as a function of the size of the system cell along the z direction. 
    
    The considered system has an extension along the vertical axis of about 11.0 Bohr. When looking at the difference in energy between the double-cell and the reference calculations, it appears that minimal artifacts are present at the smallest cell sizes (22-23 Bohr). This suggests that no significant spills in the electronic density are present for these DFT cells, consistent with our previous observations for the isolated system. As observed in the 0D case, double-cell energies are fully converged with respect to the DFT system cell already for the smallest cells considered. The results are in line with the convergence
        
    \noindent observed for a parabolic (also known as point-counter-charge or dipole correction) scheme, while AFC90  shows a slower convergence, with exact results for cells larger than 35 Bohr.

    Eventually, in the bottom panels of Figure \ref{fig:2d-res} we report the computational speedup of using a double cell approach, compared to using the same expanded cell for both electrostatic and DFT simulations. The relative weight of electrostatic and inverting the DFT Hamiltonian for this small system provide a speedup of about 2.5-2.8 in vacuum, that drops to about 1.5 in a dielectric medium. While the scaling of the FFT-based electrostatic is expected to be better than the DFT one as a function of the cell size, speedups appear to be independent of cell size for the considered system. This is probably due to the relatively small number of electrons in the system and to the other computational overheads associated with larger FFT grids. 
    
\subsection{Wire (1D) System}\label{sec:1D}

    \begin{wrapfigure}{r}{0.5\textwidth}
        \begin{center}
            \includegraphics[scale=0.125]{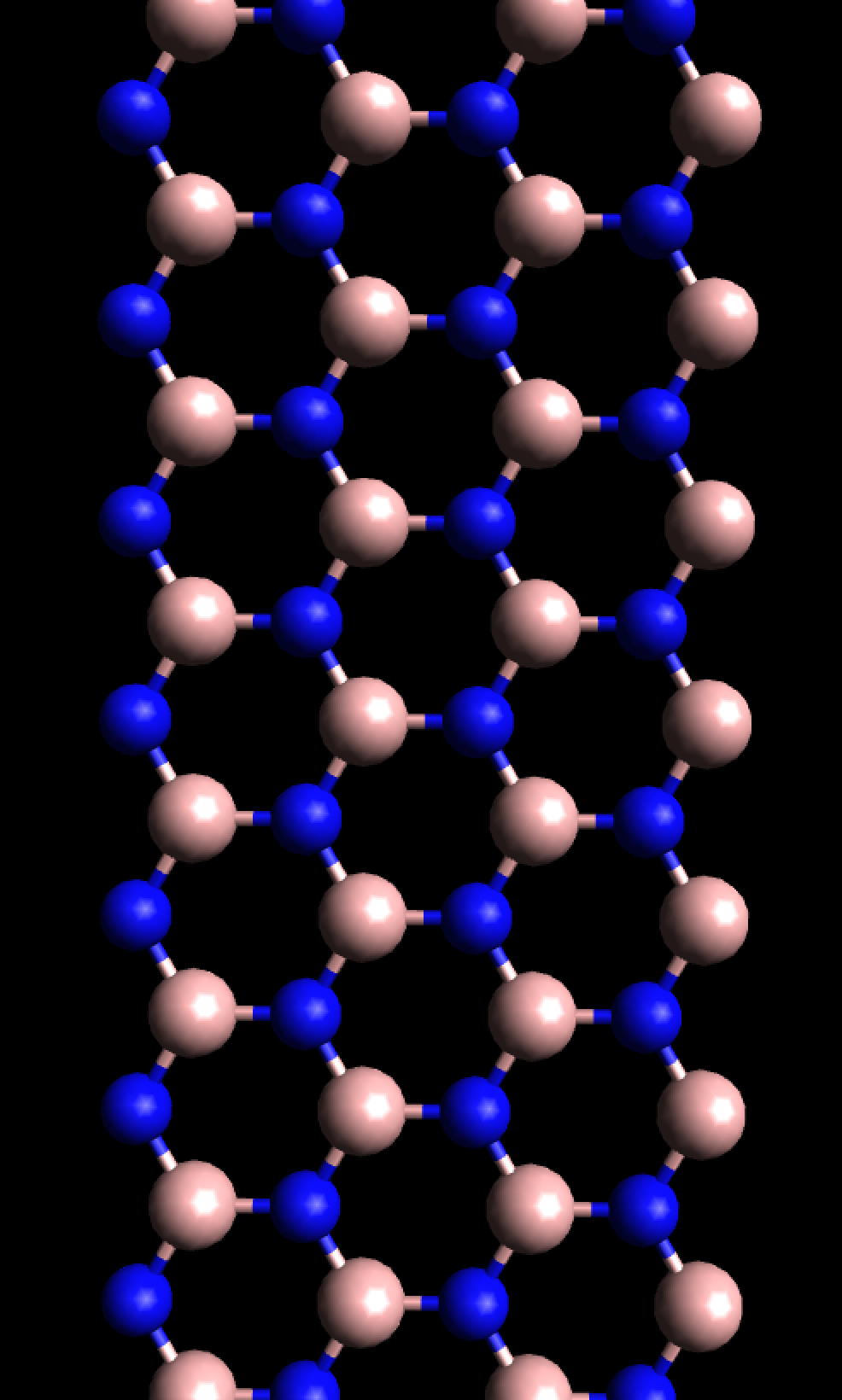}
        \end{center}
        \caption{Visualization of the BN nanoribbon (boron atoms in pink, nitrogen atoms in blue) investigated to test the double-cell formalism for one-dimensional systems.}
        \label{fig:bnrib}
    \end{wrapfigure}
    \begin{figure}[H]
        \begin{center}
            \centerline{\includegraphics[scale=0.35]{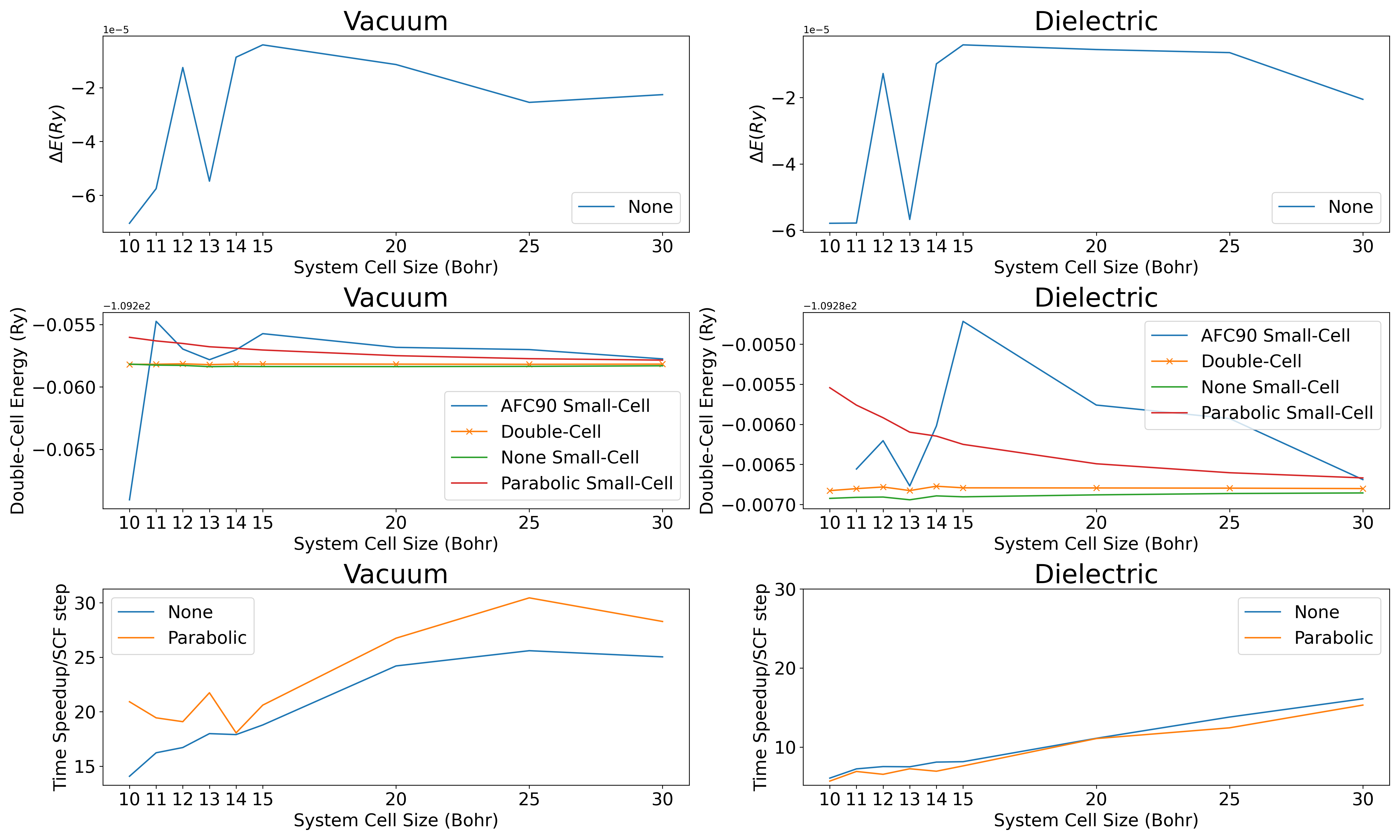}}
        \end{center}
        \caption{The top row plots show the energy difference, in Ry, between an SCF calculation using the double cell formalism and the corresponding reference calculation (as visualized in Figure 2). In the middle row, the convergence of different PBC schemes as a function of DFT system cell size is reported. The bottom row plots show the time speedup between these simulations. All reported tests were performed for systems in vacuum (left column) and in a continuum dielectric medium (right column).}
        \label{fig:1d-res}
    \end{figure}
    
    For the 1D systems, we studied the boron nitride (BN) nanoribbon, as shown in Figure \ref{fig:bnrib}. An orthorhombic cell is adopted, with the ribbon oriented along the x-axis, making the y and z axes the two non-periodic directions. Also in this case, only these two directions will be considered in the expansion of the system cell into the environment cell, while the system and environment cells have the same size along the x direction. For a sake of simplifying the discussion, the sizes of the cell axes in the non-periodic directions are kept identical, allowing to focus our analysis on a single parameter. Convergence tests are reported in Figure \ref{fig:1d-res} as a function of the such a parameter. 
    
    The BN nanoribbon has a width along the y direction of 13 Bohr, with a one-atom thickness of less than 1 Bohr in the z direction. Convergence of the double-cell calculations with respect to the reference setup show minor artifact at the smallest cells considered (10 Bohr). This is consistent with our understanding of these small-cell artifacts reported for the 0D case and with the size of the studied system. While a small drift in convergence is observed at the larger cell sizes, the associated errors are within the convergence threshold of the performed SCF calculations. 

    Similarly to the results reported for 2D systems, double-cell energies appear to be well converged at all the considered system-cell sizes. AFC90 results show a slightly slower convergence, with some fluctuations for cell sizes smaller than 20 Bohr. While the parabolic (point-counter-charge) correction is not implemented for 1D systems, in Figure \ref{fig:1d-res} we report results of using a 2D scheme applied considering the y axis as the only non-periodic direction of the system. As the system is very homogeneous along the z direction, we expect this scheme to capture most of the interaction energy between periodic replicas of the nanoribbon. Indeed, our simulations show that this parabolic correction is close to the double-cell results, although some small deviations are still present for cell sizes as large as 20 Bohr.
    
    For this application, more substantial computational speedups are observed in using the double-cell formalism. Moreover, a linear increase in the speedup is observed as a function of the cell size, consistent with the expected difference in computational scaling between the electrostatic and the DFT components of the calculations. As for the 0D and 2D cases, the speedup is significantly reduced for the simulations in a dielectric medium, as the iterations involved in the solution of the more complex Generalized Poisson problem increase the weight of the electrostatic calculation. 

\section{Conclusion}
    
    We report the development and implementation of a double-cell numerical algorithm to decouple electrostatic calculations and environment effects from the underlying DFT simulation. The proposed formalism was tested as a tool to remove PBC artifacts for partially periodic and non-periodic systems. By comparing double-cell calculations with a benchmark reference we identified a source of potential artifacts for very small system cell sizes. Indeed, if the DFT cell is so small that the electronic density of the system spills across the cell boundaries, the mapping of the DFT cell into an expanded environment cell introduces artificial cuts in the electronic density, thus forbidding the use of the double cell for minimal DFT cell sizes. However, the presented results should full convergence with cell size for most simulation setups and system dimensionalities. This fast convergence can be further improved by the additional use of PBC corrections on the environment cell. While a general speedup in simulation is observed by allowing the DFT cell to be smaller than the environment cell, the most impressive result were observed the 1D system, with speedups of up to 30 times with respect to corresponding single-cell simulations. The effectiveness and accuracy of the presented approach give us confidence in the future use of this formalism for simulations with heterogeneous and nano-scaled continuum environments. 

\bibliographystyle{unsrt}
\bibliography{main}

\begin{thebibliography}{10}
\bibitem{QMMM-rev}
Hai Lin and Donald~G. Truhlar.
\newblock Qm/mm: What have we learned, where are we, and where do we go from
  here?
\newblock {\em ChemInform}, 38(22), 2007.
\end{thebibliography}

\end{document}